\documentstyle[aps,preprint,floats,graphicx]{revtex}

\newcommand{\Ham}{{\cal H}}

\tighten
\preprint{UMIST/Phys/TP/98-2}
\begin{document}

\title{Self-consistent collective subspaces
 and diabatic/adiabatic motion in nuclei}

\author{Takashi~Nakatsukasa and Niels~R~Walet}

\address{Department of Physics, UMIST, P.O.Box 88, Manchester M60 1QD, UK}

\maketitle

\begin{abstract}
We  discuss the application of a theory
of large-amplitude collective motion to a simple model mimicking the
pairing-plus-quadrupole model of nuclear physics.
\end{abstract}

\section{Introduction}
In order to describe processes in nuclei involving large excursions
from equilibrium, such as shape-coexistence  or fission,
we need to go beyond the simple harmonic oscillator picture
underlying the random-phase approximation (RPA).
One method is to use a method that self-consistently selects
collective coordinates for  large amplitude collective motion (LACM).
Recently we have analysed properties of collective motion in a pairing model
Hamiltonian with level crossings \cite{NW98},
where we used a version of a theory proposed by Klein et al \cite{KWD91}.
to determine the collective coordinates.
We have shown that the system automatically selects
either diabatic or adiabatic collective surfaces
according to the strength of the pairing interaction.
The model used in that paper was rather artificial, since it
involved a macroscopic system coupling to a few fermionic degrees
of freedom. We felt
it  desirable to test the theory for a fully microscopic Hamiltonian
which is able to describe systems from vibrational nuclei to
well-deformed nuclei.

In this paper, we investigate the collective motion in a microscopic model
which describes a system of nucleons interacting through a simplified
version of the pairing plus quadruple force \cite{PSS80}.
Although the Hamiltonian has a very simple form,
we shall see that the model can represent
several of the interesting phenomena observed in real nuclei.

\section{Theory}

The theory of adiabatic large amplitude collective motion
(ALACM, see reference \cite{KWD91} for a complete description)
is applicable to a classical Hamiltonian system
which has kinetic terms only quadratic in momentum.
We thus have to start with a truncated Hamiltonian
\begin{equation}
\label{H_ad}
\Ham(\xi,\pi) = \frac{1}{2} 
         B^{\alpha \beta} \pi_\alpha \pi_\beta + V(\xi) \ ,
\hspace{1cm} \alpha, \beta = 1,\cdots,n \ ,
\end{equation}
where the mass tensor $B^{\alpha \beta}$, in general, depends on
the coordinates $\xi^\alpha$ and is defined by
truncation of the Hamiltonian to second order.

Collective coordinates $q^i$
and intrinsic (non-collective) coordinates $q^a$
which are approximately decoupled from each other,
are assumed to be obtainable by making a point transformation,
\begin{equation}
q^i = f^i(\xi)  \quad (i=1,\cdots, K),\qquad
q^a = f^a(\xi)  \quad (a=K+1,\cdots, n) \ .
\end{equation}
Requiring approximate decoupling between the collective and non-collective
coordinates, we find a set of coupled equations.
In the case of a single collective coordinate ($K=1$),
the equations can be written as  \cite{KWD91}
\begin{equation}
\label{LHE_1}
V_{,\alpha} = \lambda f^1_{,\alpha}, \qquad
B^{\beta\gamma} V_{;\alpha\gamma} f^1_{,\beta} = \omega^2 f^1_{,\alpha}\ .
\end{equation}

The equations (\ref{LHE_1}) can be solved
iteratively, starting from a stationary point.
In principle, the procedure to find a collective path is
to find successive points at which an eigenvector $f^1_{,\alpha}$
of the covariant RPA equation (second equation)  satisfies the force condition
(first equation) at the same time.

\section{The model}
The model Hamiltonian we discuss in this paper has
$O(4)$ symmetry.
It has been originally developed to describe $K^\pi=0^+$ excitations
in deformed nuclei \cite{PSS80}.
It has also been used to test various methods for calculation of
collective excitations \cite{Mat82,Mat86,SM88}.
The generalisation to multiple shells has also been formulated, in order
to investigate shape-coexistence phenomena \cite{FMM91}.


Consider a single shell with angular momentum $j$, and
define the four basic operators,
\begin{eqnarray}
P^\dagger &=& \sum_{m>0} c_m^\dagger c_{\bar m}^\dagger \ , \quad
\tilde{P}^\dagger = \sum_{m>0} \sigma_m c_m^\dagger c_{\bar m}^\dagger \ ,\\
N         &=& \sum_m c_m^\dagger c_m \ , \quad
Q         = \sum_m \sigma_m c_m^\dagger c_m \ .
\end{eqnarray}
Here $\sigma_m=-1$ if $|m|<\Omega/2$,  and $+1$ for the other values of $m$,
which mimics properties of the
matrix elements of the quadrupole operator $r^2 Y^2_0$.
$\Omega = j+1/2$ is assumed to be an even integer.
Hereafter we call $Q$ the quadrupole operator.
The four operators, $P$, $\tilde{P}$, $N$ and $Q$,
close under commutation, and  generate the Lie algebra $so(4)$.

We now build a model Hamiltonian
with $O(4)$ dynamical symmetry
\begin{equation}
\label{O4_Hamiltonian}
H = -G P^\dagger P - \frac{1}{2} \chi Q^2 \ ,
\end{equation}
which contains both a monopole pairing and a quadrupole-quadrupole (QQ) interaction. 
The exact solution  can be obtained by diagonalisation in an $O(4)$ basis.

The behaviour of the model is based in the competition between 
the pairing force which tends to align quasi-spins and the QQ force
which tends to de-align them. This can be made extremely clear in the
limit of no pairing or no QQ force, where the model can be solved exactly.

We can easily generalise the previous model into a multi-$j$-shell case,
$(j_1,j_2,\cdots,j_\Lambda)$.
For each $j$-shell, we take $\Omega_i = j_i + 1/2$ to be even,
and we weigh the quadrupole operators by $q_j$, the
 magnitude of quadrupole moment carried by the
single-particle state $j$.
We again use the pairing plus quadrupole Hamiltonian, but we now add
spherical single-particle energies,
\begin{equation}
H = \sum_{jm} \epsilon_j c_{jm}^\dagger c_{jm}
     -G P^\dagger P - \frac{1}{2} \chi Q^2 \ .
\end{equation}

\section{Results and Conclusion}

\begin{figure}
\includegraphics[width=\textwidth]{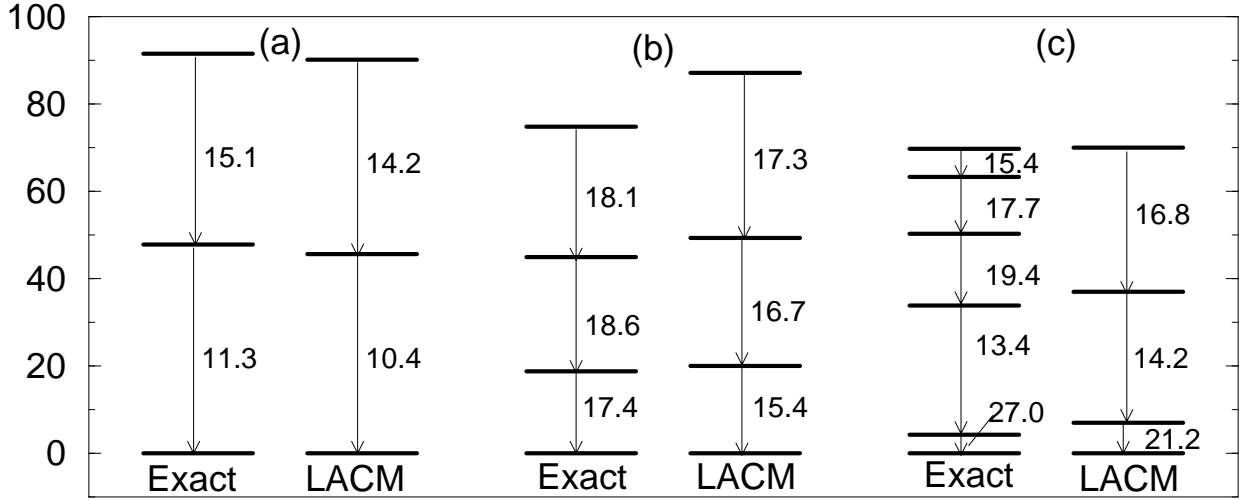}
\caption{Excitation energies  and transition matrix elements 
$|\langle n | Q | n-1 \rangle |$ (numbers next to arrows)
in the two-shell $O(4)$ model discussed in the text,
as functions of quadrupole force strength $\chi$. The case (a) represents
the almost harmonic motion for $\chi=0.1$, (b) the flat potential
for $\chi=0.3$ and (c)  shape coexistence for $\chi=0.4$.
In each case we compare an exact diagonalisation with the result
of requantisation in a single collective variable.
}
\end{figure}

One of the representative calculations we performed is for
a two-shell problem. Both shells have $\Omega=10$, but we take
$q_1=3,q_2=1$ and $\epsilon_1=0,\epsilon_2=10$. For strong ``deformation''
we expect that the shells mix strongly, and for weak QQ force we expect a
simple spectrum. In this case, as can be seen in figure 1, 
we find that
as we move from a harmonic spectrum to a more complex mixed situation,
 that we 
still keep reasonable agreement for both spectrum and transition  strengths.
The values of the QQ strength used were such that the middle case exactly
captures the turn-over from single to double well. In all these cases
one collective coordinate appears to be enough. There are some states in the
double-well structure that are not well described by the current method,
which is currently under investigation.

In conclusion we have shown that our method can deal with both anharmonic
vibrations and shape coexistence. In future we shall extend this research
to the proton-neutron O(4) model, and to the real pairing-plus-quadrupole
model.

This work is supported by  a research grant (GR/L22331) from the EPSRC.

\newcommand{\PR}{{\em Phys. Rev.} }
\newcommand{\NP}{{\em Nucl. Phys.} }
\newcommand{\APNY}{{\em Ann. Phys.} NY}
\newcommand{\PTP}{{\em Prog. Theor. Phys.} }

\end{document}